# NEUTRON COUNTING STATISTICS CALCULATIONS USING DETERMINISTIC TRANSPORT

**Philippe Humbert**

CEA, DAM, DIF, F-91297 Arpajon, France

philippe.humbert@cea.fr

## ABSTRACT

For a number of applications like low-source reactor start-up or neutron coincidence counting it is necessary to take into account the stochastic nature of neutron transport and go beyond the average neutron density, which is solution of a linear Boltzmann equation. In this work, we are particularly interested in calculating the moments and probabilities of the number of neutrons detected during a time window. It is known that in case of a single initial neutron these quantities are solution of a system of coupled adjoint transport equations and that a neutron source can be taken into account in a second time using summations with the source strength. The purpose of the present work is first to present the derivation of these equations in a form where they can be solved up an arbitrary order and where the fission terms are expanded according to the moments of the fission multiplicity. For simplicity, the derivation is first presented in a lumped also called point model approximation before considering the full phase-space case. Thereafter, we describe the implementation of the solution algorithm in the deterministic, discrete ordinates code PANDA and finally, we present some numerical results and inter-code comparisons for verification purposes and to illustrate the applicability of the method.

## I. INTRODUCTION

The behavior of the neutron population is generally described by the average neutron density, which is solution of a linear Boltzmann equation, the neutron transport equation. However, neutron production, interaction with matter and detection are fundamentally stochastic processes, the most important phenomenon being the neutron multiplication by fission, which can be seen as a branching process[1,2]. Taking into account the stochasticity of neutron transport is mandatory for some applications like reactor start-up with low source[3,4], extinction/survival probability of the neutron population[5] and coincidence counting for fissile matter assay[6].

In this work, we are interested in the simulation of neutron time correlation measurements. Most of the methods currently in use are based on the statistical analysis of the number of detections over a time interval. The most popular approaches feature the first three moments of the detection number distribution as proposed by Böhnel[7] or Hage and Ciffareli[8]. The neutron number counting probability distribution is also considered by other authors[9,10]. The aim of these passive, non-destructive techniques is to link the correlation measurements to some parameters of the fissile assembly under study, such as multiplication, source intensity and type.

To this end, lumped model, where the phase space is reduced to a single point, are popular because they provide analytical relationships. On the other hand, detailed and accurate but costly simulations are carried out using 3D, continuous energy Monte Carlo codes in analog mode[11,12]. Deterministic transport codes are also applied to solve the discrete form of probability and moment full phase-space equations. The latter are obtained from the probability generating function (PGF) equations derived independently by Pàl[13] and Bell[14] using the Backward Kolmogorov master equation formulation associated to the neutron Markov processes governing the neutron life. Consequently, the main problem is to solve a system of adjoint transport-like equations.

Several authors have presented the adaptation of deterministic transport solvers to this problem. Previously, we implemented the computation of the first two moments with the discrete ordinates transport code PANDA[15]. Mattingly[16] presented an application of the deterministic transport to the computation of neutron multiplicity statistics. Endo et al.[17] considered the deterministic calculation of the third order neutron correlation. Davis-Fichtl et al.[18,19] have implemented the first four moments calculation in the deterministic transport code PARTISN[20]. Saxby et al.[21] have investigated the numerical solution of the neutron number probabilities and moments up to an arbitrary higher order using an explicit calculation of the coupling term. Pazsit and Pál[22] also considered multiplicity counting solutions beyond the point model.

Our objective is to present the numerical solution of the moments and probabilities of the detection number up to an arbitrary order using a deterministic transport code. In the following, section II details the derivation of the corresponding equations under the point model approximation using the Pàl and Bell methodology. Thus, coupled adjoint differential equations are obtained for the case of a distribution induced by one initial neutron and, subsequently, a neutron source is accounted for. The fission terms, which appear in the equations, are developed according to the moments of the fission multiplicity, so that approximations can be obtained by truncated this expansion. Thereafter, section III presents the generalization to the full neutron-phase space, resulting in adjoint transport equations. We will also examine the adaptation of detection equations to the population and leakage problem. Finally, section IV describes the implementation using the deterministic discrete ordinates code PANDA[23] and two test problems are defined to verify the applicability of the code.

## II. POINT MODEL EQUATIONS

Before considering the full phase-space problem, the derivations are performed within a lumped model approximation. In this model, we consider the time evolution of monokinetic neutrons traveling in an infinite homogeneous medium. The neutron population and the number of detections evolve without memory, which is characteristic of Markov processes. The detection number probabilities obey a master equation, the Chapmann-Kolmogorov equation. We will follow the Pàl-Bell methodology based on the backward probability generating function (PGF). Probabilities and moments are derived from the PGF, and a neutron source is taken into account in a second step.

### II.A. Single Initial Neutron Case

*II.A.1. Probability Balance and Probability Generating Function Equations*

II.A.1.a. Probability balance

The quantity of interest is the number of detections at a final time $T$ induced by one single neutron at an initial time $t$. More generally, the probability of $n$ detections with $i$ initial neutrons is written as:

$$p_n^{(i)}(t) = p(n,T|i,t) \qquad (1)$$

$$p_n(t) = p_n^{(1)}(t) \qquad (2)$$

A fraction $\varepsilon_c$ of the neutrons captured during the measurement time gate $[0,T]$ are detected. The time dependent detection efficiency writes:

$$\varepsilon_c(t) = \varepsilon_c \Theta(T-t)\Theta(t) \qquad (3)$$

Where $\Theta(t)$ is the Heaviside step function,

In the point model approximation, two reactions are considered, fission and capture, with capture representing all non-fission events, including leakage. The neutron velocity is $v$ and the probability of interaction by unit distance are the macroscopic cross-sections for fission ($\sigma_f$) and capture ($\sigma_c$), the total absorption cross section is sum of fission and capture $(\sigma_a = \sigma_f + \sigma_c)$. An induced fission can produce $i$ neutrons with the probability $f_i$. The maximum number of emitted neutrons by one fission is supposed to be lower or equal to $I$.

A differential equation for $p_n$ is obtained by considering all the possible mutually exclusive events (nothing, fission and capture with or without detection) that can arise to one neutron during the infinitesimal time interval [t-dt,t]:

$$-\frac{1}{v}\frac{dp_n}{dt} + \sigma_a p_n = \sigma_f \sum_{i=0}^{I} f_i p_n^{(i)} + \sigma_c\big(\varepsilon_c(t)\delta_{n,1} + (1-\varepsilon_c(t))\delta_{n,0}\big) \qquad (4)$$

$$p_n(T) = \delta_{n,0}$$

II.A.1.b. Probability generating function g(x)

In order to handle with the $p_n^{(i)}$ term and to derive the moment equations it is useful to introduce the PGF defined by:

$$g(x,t) = \sum_{n=0}^{\infty} x^n p_n(t) \qquad (5)$$

Where $x$ is a real number between zero and one. The generating function has the following useful property:

$$g^{(i)} = \sum_{n=0}^{\infty} x^n p_n^{(i)} = \left(\sum_{n=0}^{\infty} x^n p_n\right)^i \qquad (6)$$

Using (5) and (6) in (4), we obtain the Pàl-Bell equation for the PGF:

$$-\frac{1}{v}\frac{\partial g}{\partial t} + \sigma_a g = \sigma_f \sum_{i=0}^{I} f_i g^i + \sigma_c\big(1 - \varepsilon_c(t)(1-x)\big) \text{ with } g(x,T) = 1 \qquad (7)$$

The derivatives of the generating function give the probabilities $p_n$ and binomial moments $m_n$:

$$p_n = \frac{1}{n!}\left(\frac{\partial^n g}{\partial x^n}\right)_{x=0} \quad ; \quad m_n = \frac{1}{n!}\left(\frac{\partial^n g}{\partial x^n}\right)_{x=1} \qquad (8)$$

The binomial moment also called reduced factorial or combinatorial moment is the mean value of the binomial coefficient $\binom{k}{n}$, the number of combinations of $k$ elements among $n$.

$$m_k = \sum_{n=k}^{\infty} \binom{n}{k} p_n \qquad (9)$$

II.A.1.c. Generating function h(x)

Considering the fission term in equation (7), it is customary to work with the binomial moments of the fission multiplicity $v_i$ instead of the probabilities $f_i$. In this way, approximations are obtained by truncating the fission term. For example, in the quadratic approximation, moments of order greater than two are neglected. To this end, we introduce the generation of the function $h(x,t)$ defined by:

$$h(x,t) = 1 - g(x,t) \tag{10}$$

We introduce the adjoint operator $L^+$, which accounts for adjoint time dependency and absorption plus fission, which appear in the conventional adjoint neutron density equation. This operator will be generalized in section III to take into account the full phase-space dependency.

$$L^+ = -\frac{1}{v}\frac{\partial}{\partial t} + (\sigma_a - \bar{v}\sigma_f) \tag{11}$$

Using (10) and (11) in (7), we obtain the equation for $h(x,t)$:

$$L^+ h = -\sigma_f \sum_{i=2}^{I} v_i(-h)^i + \sigma_c \varepsilon_c(t)(1-x) \quad \text{with } h(x,T) = 0 \tag{12}$$

Counting probabilities and moments are derived from $h(x)$ using the following relationships:

$$h(x=0) = 1 - p_0 = \pi_0 \quad ; \quad p_n = \frac{-1}{n!}\left(\frac{\partial^n h}{\partial x^n}\right)_{x=0} \tag{13}$$

$$h(x=1) = 0 \quad ; \quad m_n = \frac{-1}{n!}\left(\frac{\partial^n h}{\partial x^n}\right)_{x=1} \tag{14}$$

$\pi_0$ is the probability of counting more than zero neutron.

*II.A.2. Equation of probabilities*

II.A.2.a. Equation for $\pi_0$

Setting $x$ to zero in equation (12) gives:

$$L^+ \pi_0 = -\sigma_f \left(\sum_{i=2}^{I} v_i(-\pi_0)^i\right) + \sigma_c \varepsilon_c(t) \quad \text{with } \pi_0(T) = 0 \tag{15}$$

We note that the probability of counting more than zero neutrons during the time gate is solution of a non-linear equation.

II.A.2.b. Equation for $p_1$

According to equation (14), the probability $p_1$ is given by:

$$p_1 = -\left(\frac{\partial h}{\partial x}\right)_{x=0} \tag{16}$$

Furthermore, we note that:

$$\left(\frac{\partial h^i}{\partial x}\right)_{x=0} = -i\pi_0^{i-1} p_1 \qquad (17)$$

Using (16) and (17) in (12), we obtain the equation for $p_1$:

$$L^+ p_1 = \sigma_f \left(\sum_{i=2}^{I} i v_i (-\pi_0)^{i-1}\right) p_1 + \sigma_c \varepsilon_c(t) \text{ with } p_1(T) = 0 \qquad (18)$$

The equation for the probability of one count during the time gate is solution of a linear equation with an additional fission term, which depends on the probability $\pi_0$.

II.A.2.c. Equation for $p_n$

The equation for $p_n$, with $n > 1$, is obtained by using (13) in equation (12). We introduce the following notation:

$$\pi_n^{(i)} = \frac{(-1)^i}{n!} \left(\frac{\partial^n h^i}{\partial x^n}\right)_{x=0} \qquad (19)$$

In particular, we have:

$$\pi_n^{(1)} = p_n \qquad (20)$$

$$\pi_0^{(1)} = -\pi_0 = p_0 - 1 \qquad (21)$$

$$\pi_0^{(i)} = (-1)^i \pi_0^i \qquad (22)$$

Using definition (19) in equation (12), we obtain:

$$L^+ p_n = \sigma_f \sum_{i=2}^{I} v_i \pi_n^{(i)} \qquad (23)$$

To expand $\pi_n^{(i)}$, we apply the Leibniz formula, which states that, the nth-order derivative of $h^i$ is:

$$\left(\frac{\partial^n h^i}{\partial x^n}\right) = \sum_{k=0}^{n} \binom{n}{k} \left(\frac{\partial^{n-k} h^{i-1}}{\partial x^{n-k}}\right) \left(\frac{\partial^k h}{\partial x^k}\right) \qquad (24)$$

Using notation (19) in (24), we obtain the following relation, which will be useful for the recursive calculation of $\pi_n^{(i)}$:

$$\pi_n^{(i)} = \sum_{k=0}^{n} \pi_{n-k}^{(i-1)} \pi_k^{(1)} \qquad (25)$$

We denote $\pi_n^{*(i)}$ the part of $\pi_n^{(i)}$ that is independent of $p_n$. Expanding equation (25), we obtain:

$$\pi_n^{(i)} = i(-\pi_0)^{i-1} p_n + \pi_n^{*(i)} \qquad (26)$$

By introducing relation (26) in equation (23), we see that the probability $p_n$ is solution of:

$$L^+ p_n = \sigma_f \left( \sum_{i=2}^{I} i v_i (-\pi_0)^{i-1} \right) p_n + \sigma_f \sum_{i=2}^{I} v_i \pi^{*(i)}_n \quad \text{with} \quad p_n(T) = 0 \tag{27}$$

Where $\pi^{*(i)}_n$ is calculated using the following recurrence deduced from (25):

$$\pi^{*(1)}_n = 0 \quad ; \quad \pi^{*(i)}_n = -\pi^{*(i-1)}_n \pi_0 + \sum_{k=1}^{n-1} \pi^{(i-1)}_{n-k} p_k \tag{28}$$

We observe that $p_n$ is the solution of an equation similar to $p_1$, except for the source term, which can be calculated from the probabilities previously obtained using (28).

*II.A.3. Equation of moments*

Moments are calculated in the same way as probabilities, but more easily, since the zeroth-order moment is equal to zero.

II.A.3.a. Equation for $m_1$

According to equation (14), the first moment is given by:

$$m_1 = -\left( \frac{\partial h}{\partial x} \right)_{x=1} \tag{29}$$

We note that:

$$\left( \frac{\partial h^i}{\partial x} \right)_{x=1} = -i(h(1))^{i-1} m_1 = -m_1 \delta_{i,1} \tag{30}$$

Using (29) and (30) in (12), we obtain the equation for $m_1$:

$$L^+ m_1 = \sigma_c \varepsilon_c(t) \quad \text{with} \quad m_1(T) = 0 \tag{31}$$

This is the classical linear equation for the average number of detections in adjoint form.

II.A.3.b. Equation for $m_n$

The equation for $m_n$, with $n > 1$, is obtained by using (14) in equation (12). We introduce the following notation:

$$\mu^{(i)}_n = \frac{(-1)^i}{n!} \left( \frac{\partial^n h^i}{\partial x^n} \right)_{x=1} \tag{32}$$

In particular, we have:

$$\mu^{(1)}_0 = 0 \quad ; \quad \mu^{(1)}_{n>0} = m_n \quad ; \quad \mu^{(i)}_i = m_1^i \quad ; \quad \mu^{(i>n)}_n = 0 \tag{33}$$

Using the notation (32) and properties (33) in Leibniz recurrence (24), we obtain the following relation, which allows for the recursive calculation of $\mu_n^{(i)}$:

$$\mu_n^{(i)} = \sum_{k=1}^{n-1} \mu_{n-k}^{(i-1)} m_k \qquad (34)$$

Applying definition (32) in equation (12), we find the equation for the higher order moments $m_n$:

$$L^+ m_n = \sigma_f \sum_{i=2}^{n} \nu_i \mu_n^{(i)} \quad \text{with} \quad m_n(T) = 0 \qquad (35)$$

We see that $m_n$ is the solution to an equation similar to $m_1$, with the exception of the source term derived from the previously calculated moments using relation (34). It is also be noted that, unlike probabilities, the nth-order moment depends on fission multiplicity moments of order less than or equal to $n$, but not greater.

## II.B. Neutron Source

Up to now, we have been interested in the counting distribution induced by a single initial neutron. In what follows, we will consider the distribution induced by a compound Poisson neutron source. In this case, the source events are Poissonian, the distribution of the source decay time is homogeneous in time, and the number of neutrons produced by an event is a random number. A spontaneous fission source is a typical type of compound Poisson source.

### II.B.1. One initial source event

We begin by looking at the number of detections induced by one initial source event. We denote $\hat{f}_i$ the probability that $i$ neutrons are produced by one source event and $\hat{\nu}_i$ the binomial moment of order $i$ of this distribution. As in the case of induced fission, the maximum number of neutrons emitted by one source event is denoted $I$. The probability to count $n$ detections due to one initial source event is:

$$\hat{p}_n = \sum_{i=0}^{I} \hat{f}_i\, p_n^{(i)} \qquad (39)$$

The associated generating function is:

$$\hat{g}(x) = \sum_{n=0}^{\infty} x^n\, \hat{p}_n \qquad (40)$$

The relation to the function $h(x)$ is as follows:

$$\hat{g}(x) = \sum_{i=0}^{I} \hat{f}_i\, g^i = \sum_{i=0}^{I} \hat{\nu}_i\, (-h)^i = 1 + \sum_{i=1}^{I} \hat{\nu}_i\, (-h)^i \qquad (41)$$

We define the function $\hat{h}(x)$ as:

$$\hat{h}(x) = 1 - \hat{g}(x) = -\sum_{i=1}^{I} \hat{v}_i (-h)^i \tag{42}$$

The probability $\hat{p}_n$ and moment $\hat{m}_n$ of the distribution induced by one source event are expressed in terms of the $\hat{f}_i$ and $\hat{v}_i$ as follows:

$$\hat{p}_n = \sum_{i=0}^{I} \hat{f}_i p_n^{(i)} = \sum_{i=1}^{I} \hat{v}_i \pi_n^{(i)} \tag{43}$$

$$\hat{m}_n = \sum_{i=0}^{I} \hat{f}_i m_n^{(i)} = \sum_{i=1}^{n} \hat{v}_i \mu_n^{(i)} \tag{44}$$

*II.B.2. Probability equation with source*

The probability of counting $n$ neutrons at final time $T$ in presence of a neutron source between $t$ and $T$, with zero neutron present at initial time t is noted:

$$P_n = p_S(n, T | 0, t) \tag{45}$$

During an infinitesimal initial time interval $[t - dt, t]$, there are two possible mutually exclusive events producing $n$ detections at final time $T$. The first possibility is no source event during the initial time interval followed by $n$ detections between $t$ and $T$. The second possibility is one source event during the time interval, producing $i$ neutrons, followed by $n$-$i$ detections between $t$ and $T$. Consequently, we have the following equation for $P_n$:

$$-\frac{dP_n}{dt} = S\left(\sum_{i=0}^{I} P_{n-i} \hat{p}_i - P_n\right) \quad \text{with} \quad P_n(T) = \delta_{n,0} \tag{46}$$

*II.B.3. Probability generating function equation*

The generating function associated to $P_n$ is:

$$G(x) = \sum_{n=0}^{\infty} x^n P_n \tag{47}$$

Applying the definition (47) in (46), we have:

$$-\frac{\partial G(x,t)}{\partial t} = S(t) G(x,t)(\hat{g}(x,t) - 1) \quad \text{with} \quad G(x,T) = 1 \tag{48}$$

The solution is:

$$G(x,t) = \exp\left(-\int_t^T S(\tau) \hat{h}(x,\tau) d\tau\right) \tag{49}$$

*II.B.4. Probability and Moment equations*

II.B.4.a. $P_0$ expression

By definition of the PGF (46), we have:

$$P_0(t) = G(0,t) \tag{50}$$

The expression for $P_0$ is obtained by setting $x$ to zero in equation (49):

$$P_0(t) = \exp\left(-\int_t^T S(\tau)\hat{\pi}_0(\tau)d\tau\right) \tag{51}$$

$$\hat{\pi}_0 = -\sum_{i=1}^{I} \hat{v}_i (-\pi_0)^i \tag{52}$$

The probability of zero detection in the presence of a source is then:

$$P_0(t) = \exp\left(\int_t^T S(\tau) \sum_{i=1}^{I} \hat{v}_i (-\pi_0)^i d\tau\right) \tag{53}$$

In the case of an uncorrelated source (Poisson), each source event produces exactly one neutron and the probability $P_0$ is simplified as follows:

$$P_0(t) = \exp\left(-\int_t^T S(\tau)\pi_0(\tau)d\tau\right) \tag{54}$$

II.B.4.b. Calculation of $P_n$

We introduce the binomial cumulant generating function $K(x)$ defined by:

$$G(x,t) = \exp(K(x,t)) \tag{55}$$

$$K(x,t) = -\int_t^T S(\tau)\hat{h}(x,\tau)d\tau \tag{56}$$

The counting distribution is a compound Poisson distribution, with parameters $\Lambda_n$:

$$\Lambda_n(t) = \frac{1}{n!}\left(\frac{\partial^n K}{\partial x^n}\right)_{x=0} \tag{57}$$

The expression of these parameters is obtained from equations (56) and (57):

$$\Lambda_n(t) = \int_t^T S(\tau)\hat{p}_n(\tau)d\tau \tag{58}$$

$$\Lambda_n(t) = \int_t^T S(\tau)\sum_{i=1}^I \hat{v}_i \pi_n^{(i)}(\tau)\,d\tau = \int_t^T S(\tau)\sum_{i=1}^I \hat{f}_i p_n^{(i)}(\tau)\,d\tau \tag{59}$$

In case of an uncorrelated source, the parameters are:

$$\Lambda_n(t) = \int_t^T S(\tau)p_n(\tau)d\tau \tag{60}$$

Probabilities are calculated using the parameters of the compound Poisson distribution in a recurrence that has been established by Panjer[24]. This recurrence states that:

$$nP_n(t) = \sum_{i=1}^n i\Lambda_i(t)P_{n-i}(t) \tag{61}$$

Hage and Cifarelli[8] also used this formula in the context of neutron multiplicity counting.

II.B.4.c. Calculation of $M_n$

As noted above, the binomial moment calculation is similar to the probability case, but simpler, since we know that the moment of order zero is equal to one.

The binomial cumulant $\Gamma_n$ are the derivatives of the cumulant generating function.

$$\Gamma_n(t) = \frac{1}{n!}\left(\frac{\partial^n K}{\partial x^n}\right)_{x=1} \tag{62}$$

The expression of $\Gamma_n$ is obtained by means of equations (56) and (62):

$$\Gamma_n(t) = \int_t^T S(\tau)\hat{m}_n(\tau)d\tau \tag{63}$$

$$\Gamma_n(t) = \int_t^T S(\tau)\sum_{i=1}^I \hat{v}_i \mu_n^{(i)}(\tau)\,d\tau = \int_t^T S(\tau)\sum_{i=1}^I \hat{f}_i m_n^{(i)}(\tau)\,d\tau \tag{64}$$

The first binomial cumulant is the average number of counts, the second binomial cumulant is the average number of correlated double counts and more generally, $\Gamma_n$ is the average number of combinations of *n* correlated counts (n-uple).

The Feynman relative moments $Y_n$ are often used in multiplicity counting applications, they are written in terms of binomial cumulants as:

$$Y_n = \frac{n!\, \Gamma_n}{\Gamma_1} \tag{65}$$

In case of an uncorrelated source, the binomial cumulant expression (64) simplifies to:

$$\Gamma_n(t) = \int_t^T S(\tau) m_n(\tau) d\tau \tag{66}$$

Like probabilities, binomial moments are calculated using the binomial cumulants in Panjer's recurrence applied to moments:

$$nM_n(t) = \sum_{i=1}^{n} i\Gamma_i(t) M_{n-i}(t) \tag{67}$$

## III. GENERAL EQUATIONS

### III.A. Adjoint Transport Equation

The stochastic neutronics, point model equations studied in the previous section can be generalized to take into account the complete phase space, characterized by neutron position $(\vec{r})$, direction $(\vec{\Omega})$ and kinetic energy $(E)$.

The adjoint operator of the point model (11) becomes the operator of the conventional adjoint neutron transport equation, which includes advection, diffusion and induced fission. The generalized $L^+$ operator applied to the adjoint function $\varphi+$ is as follows:

$$L^+ \varphi^+(\vec{r}, E, \vec{\Omega}, t) = \left( -\frac{1}{v(E)} \frac{\partial}{\partial t} - \vec{\Omega} \cdot \vec{\nabla} + \sigma_T(\vec{r}, E) - S^+ - F^+ \right) \varphi^+(\vec{r}, E, \vec{\Omega}, t) \tag{68}$$

In which $S^+$ and $F^+$ are the adjoint scattering and fission operators defined by:

$$S^+ \varphi^+(\vec{r}, E, \vec{\Omega}, t) = \sigma_S(\vec{r}, E) \iint dE' d^2\Omega' f_S(\vec{r}, E \to E', \vec{\Omega}.\vec{\Omega}') \varphi^+(\vec{r}, E', \vec{\Omega}', t)$$

$$F^+ \varphi^+(\vec{r}, E, \vec{\Omega}, t) = \bar{\nu}(\vec{r}, E) \sigma_F(\vec{r}, E) \int dE' \chi(\vec{r}, E, E') \int d^2\Omega' \, \varphi^+(\vec{r}, E', \vec{\Omega}', t)$$

Where $f_S(\vec{r}, E \to E', \vec{\Omega}.\vec{\Omega}')$ is the scattering kernel and $\chi(\vec{r}, E, E')$ is the fission spectrum induced by a neutron of energy $E$.

The adjoint transport equation with adjoint source term $Q^+$, and zero final time and outgoing boundary condition conditions are:

$$L^+ \varphi^+(\vec{r}, E, \vec{\Omega}, t) = Q^+(\vec{r}, E, \vec{\Omega}, t) \tag{69}$$

$$\varphi^+(\vec{r}, E, \vec{\Omega}, T) = 0 \tag{70}$$

$$\varphi^+(\vec{r}, E, \vec{\Omega}, t) = 0 \quad \vec{n}.\vec{\Omega} > 0 \quad \vec{r} \in \partial D \tag{71}$$

Where $\partial D$ is the external boundary surface and $\vec{n}$ is the outer unit vector normal to $\partial D$.

In the following, the adjoint functions averaged over the induced fission spectrum $\chi(\vec{r}, E, E')$ and source spectrum $\hat{\chi}(\vec{r}, E')$ will be noted:

$$\overline{\varphi}^+(\vec{r}, E, t) = \int dE' \chi(\vec{r}, E, E') \int d^2\Omega' \varphi^+(\vec{r}, E', \vec{\Omega}', t) \tag{72}$$

$$\tilde{\varphi}^+(\vec{r}, E, t) = \int dE' \hat{\chi}(\vec{r}, E') \int d^2\Omega' \varphi^+(\vec{r}, E', \vec{\Omega}', t) \tag{73}$$

Using notation (72), the adjoint induced fission term is written as:

$$F^+ \varphi^+(\vec{r}, E, \vec{\Omega}, t) = \overline{\nu}\sigma_F(\vec{r}, E)\overline{\varphi}^+(\vec{r}, E, t) \tag{74}$$

**III.B. Probability and Moment Equations**

*III.B.1. One single initial neutron*

III.B.1.a. Non-linear adjoint transport equation for $\pi_0$

The full phase space extension of point model equation (15) for $\pi_0$ is:

$$L^+ \pi_0(\vec{r}, E, \vec{\Omega}, t) = -\sigma_F(\vec{r}, E) \sum_{i=2}^{I} \nu_i(\vec{r}, E)\left(-\overline{\pi}_0(\vec{r}, E, t)\right)^i + \sigma_C(\vec{r}, E)\varepsilon_C(\vec{r}, t) \tag{75}$$

In addition to the adjoint transport operator, there is a supplementary non-linear fission operator $F_{NL}^+$:

$$F_{NL}^+ \cdot \pi_0(\vec{r}, E, \vec{\Omega}, t) = -\sigma_F(\vec{r}, E) \sum_{i=2}^{I} \nu_i(\vec{r}, E)\left(-\overline{\pi}_0(\vec{r}, E, t)\right)^i \tag{76}$$

The adjoint source tem is:

$$Q_C^+(\vec{r}, E, t) = \sigma_C(\vec{r}, E)\varepsilon_C(\vec{r}, t) \tag{77}$$

Therefore, we can rewrite equation (75) as follows:

$$(L^+ - F_{NL}^+) \cdot \pi_0(\vec{r}, E, \vec{\Omega}, t) = Q_C^+(\vec{r}, E, t) \tag{78}$$

III.B.1.b. Adjoint transport equation for $p_1$

The generalization of point model equation (18) for $p_1$ is:

$$L^+ p_1(\vec{r}, E, \vec{\Omega}, t) = \sigma_F(\vec{r}, E)\left(\sum_{i=2}^{I} i\nu_i(\vec{r}, E)\left(-\overline{\pi}_0(\vec{r}, E, t)\right)^{i-1}\right)\overline{p}_1(\vec{r}, E) +$$
$$+\sigma_C(\vec{r}, E)\varepsilon_C(\vec{r}, t) \tag{79}$$

With respect to the conventional transport equations, there is an additional linear fission operator given by:

$$F_L^+ \cdot p_1(\vec{r}, E, \vec{\Omega}, t) = \sigma_F(\vec{r}, E) \left( \sum_{i=2}^{I} i v_i(\vec{r}, E) \left(-\overline{\pi}_0(\vec{r}, E, t)\right)^{i-1} \right) \bar{p}_1(\vec{r}, E, \vec{\Omega}, t) \quad (80)$$

Therefore, we can rewrite equation (79) as:

$$(L^+ - F_L^+) \cdot p_1(\vec{r}, E, \vec{\Omega}, t) = Q_C^+(\vec{r}, E, t) \quad (81)$$

III.B.1.c. Adjoint transport equation for $p_n$

The generalization of point model equation (27) for $p_n$, with n > 1 is:

$$L^+ p_n(\vec{r}, E, \vec{\Omega}, t) = \sigma_F(\vec{r}, E) \left( \sum_{i=2}^{I} i v_i(\vec{r}, E) \left(-\overline{\pi}_0(\vec{r}, E, t)\right)^{i-1} \right) \bar{p}_n(\vec{r}, E) + \\ + \sigma_F(\vec{r}, E) \sum_{i=2}^{I} v_i(\vec{r}, E) \overline{\pi}_n^{*(i)}(\vec{r}, E) \quad (82)$$

In this case, the adjoint source term is:

$$Q_{P,n}^+(\vec{r}, E, t) = \sigma_F(\vec{r}, E) \sum_{i=2}^{I} v_i(\vec{r}, E) \overline{\pi}_n^{*(i)}(\vec{r}, E, t) \quad (83)$$

This source term is calculated from the previously obtained probabilities using recurrence (28). We can rewrite equation (82) as below:

$$(L^+ - F_L^+) \cdot p_n(\vec{r}, E, \vec{\Omega}, t) = Q_{P,n}^+(\vec{r}, E, t) \quad (84)$$

III.B.1.d. Adjoint transport equation for the binomial moments

The full phase-space equivalent of the point model equations (31) and (35) are:

$$L^+ m_1(\vec{r}, E, \vec{\Omega}, t) = \sigma_C(\vec{r}, E) \varepsilon_C(\vec{r}, t) \quad (85)$$

$$L^+ m_n(\vec{r}, E, \vec{\Omega}, t) = \sigma_F(\vec{r}, E) \sum_{i=2}^{n} v_i(\vec{r}, E) \overline{\mu}_n^{(i)}(\vec{r}, E) \quad (86)$$

These are conventional adjoint transport equations with adjoint source terms $Q_C^+$ defined in (77) and $Q_{M,n}^+$ given by:

$$Q_{M,n}^+(\vec{r}, E, t) = \sigma_F(\vec{r}, E) \sum_{i=2}^{I} v_i(\vec{r}, E) \overline{\mu}_n^{(i)}(\vec{r}, E, t) \quad (87)$$

This source term is calculated from the previously obtained probabilities using recurrence (34).

We can rewrite equation (85) and (86) as:

$$L^+ \cdot m_1(\vec{r}, E, \vec{\Omega}, t) = Q_C^+(\vec{r}, E, t) \quad (88)$$

$$L^+ \cdot m_n(\vec{r}, E, \vec{\Omega}, t) = Q_{M,n}^+(\vec{r}, E, t) \quad (89)$$

The boundary and final conditions associated with the moment and probability equations are zero final time and the outgoing boundary condition given in (70) and (71).

*III.B.2. Neutron number and leakage*

Similar equations are obtained when we consider the number of neutrons present instead of the number of neutrons detected. To do this, simply remove the detection term ($\varepsilon_C = 0$) and modify the final condition so that one neutron is present at final time:

$$m_n(T|\vec{r}, E, \vec{\Omega}, T) = p_n(T|\vec{r}, E, \vec{\Omega}, T) = \delta_{n,1} \quad (90)$$

If we are interested in the number of particles escaping from the system during the time interval $[0, T]$ we need only remove the detection term ($\varepsilon_C = 0$) and change the boundary conditions to:

$$m_n(T|\vec{r}, E, \vec{\Omega}, t) = p_n(T|\vec{r}, E, \vec{\Omega}, t) = \Theta(t)\Theta(T-t)\delta_{n,1} \text{ when } \vec{r} \in \partial D \text{ and } \vec{n}.\vec{\Omega} \geq 0 \quad (91)$$

Where $\Theta(t)$ is the Heaviside step function.

*III.B.3. Neutron source*

The point model equation (53) for the zero-count probability in the presence of a neutron source is generalized as follows:

$$P_0(t) = \exp\left(\int_t^T d\tau \int d^3r \, S(\vec{r}, \tau) \sum_{i=1}^{I} \hat{v}_i(\vec{r})\left(-\tilde{\pi}_0(\vec{r}, t)\right)^i\right) \quad (92)$$

The parameters of the compound Poisson counting distribution are:

$$\Lambda_n(t) = \int_t^T d\tau \int d^3r \, S(\vec{r}, \tau) \sum_{i=0}^{I} \hat{v}_i(\vec{r})\tilde{\pi}_n^{(i)}(\vec{r}, t) \quad (93)$$

Thereafter, the probabilities are successively calculated using Panjer's recurrence (61).

Likewise, the generalization of the point model binomial cumulant expression (64) is:

$$\Gamma_n(t) = \int_t^T \int d^3r \, S(\vec{r}, \tau) \sum_{i=1}^{n} \hat{v}_i(\vec{r})\tilde{\mu}_n^{(i)}(\vec{r}, t) \, d\tau \quad (94)$$

The binomial moments are calculated successively using Panjer's recurrence applied to the moments (67).

## IV. NUMERICAL SOLUTION AND TEST PROBLEMS

In the previous section, we saw that the stochastic neutronics equations with a single initial neutron and no source are coupled adjoint transport equations. They can therefore be solved using conventional deterministic transport codes with a minimum of adaptation work. In this study, we consider the numerical solution in full phase space using the PANDA neutral particles transport code.

### IV.A. Numerical Solution Using PANDA Code

*IV.A.1. Adjoint calculation with PANDA*

PANDA code is based on conventional discretization and solution methods. It is multigroup in energy, discrete ordinates in angle, weighted-diamond in space and an implicit or semi-implicit Crank-Nicholson scheme is used for time discretization. Scattering anisotropy is taken into account by means of a Legendre polynomial expansion of the scattering term. The system of multigroup equations is solved by classical dependent source iteration method. The transport code is adapted for adjoint calculations by reversing the energy group order and transposing the transfer cross section matrices. Both stationary and time-dependent calculations can be carried out. Time-dependent calculations begin with the final condition and proceed backwards until the initial time is reached. Stationary calculations are performed when the time derivative is set to zero in (68).

*IV.A.2. Calculation of moments*

The moment equations form a system of linear time-dependent adjoint transport equations. At each time step, all the binomial moments are calculated sequentially, in increasing order, starting from the first moment, the average count number, solution of (85). The subsequent moment equations (86) depend on all previously determined moments via a recursively calculated adjoint source term (87) by means of (34).

Afterward, the external source is taken into account to update the binomial cumulants $\Gamma_n$ using (94) and the binomial moments $M_n$ are computed by means of the Panjer's recurrence (67). The time dependent adjoint calculation ends when the contribution of one neutron born before the detector opens is negligible.

*IV.A.3. Calculation of probabilities*

A similar procedure is applied for calculating the probabilities. However, it is more complex because, unlike the zero-order moment, which is simply equal to unity, the zero-count probability $p_0$ (or the probability $\pi_0$ of the complementary event) is solution of a non-linear adjoint transport equation (75). In fact, this non-linearity problem can be solved without modifying the standard algorithm by using source iterations whose number is slightly increased. The subsequent equations for counting one or more neutrons remain linear, but include an additional linear induced fission source term (76).

After calculating the initial single neutron probability $p_n$ on the current time-step, the external source is taken into account and the compound Poisson distribution parameter $\Lambda_n$ is updated by integration with the neutron source over the time step using (93). Then, the counting probability, with source, $P_n$ is derived using the Panjer's recurrence (61). If the parameter $\Lambda_n$ becomes negligible for $n \geq N$, it can be set to zero in the Panjer's recurrence, and the corresponding adjoint transport equations do not need to be solved, thus reducing the computational burden.

**IV.B. Test problems**

*IV.B.1. First test problem*

In order to verify the implementation of these algorithms in the PANDA discrete ordinates code, we present some comparisons between the results obtained with the reference Monte Carlo MCNP6.1[25] on a test problem.

IV.B.1.a. Test Problem description

The 1D spherical test problem is presented in Figure 1 and Table I. It consists of a highly enriched uranyl nitrate solution with an internal BF3 detector and a $^{252}$Cf spontaneous fission source placed on the external boundary surface. The source intensity is 20,000 neutrons per second. The detection number is the number neutron captured by the BF3 detector.

The relevant quantities are the stationary and time dependent count number probability distributions, along with the corresponding first moments (mean count rate and Feynman moments). The Feynman moment $Y_n$ is related to the correlated combinatorial moment $\Gamma_n$, and is defined in (65).

**Table I.** Characteristics of test problem 1.

| Shell | Material | Isotope | at% | $R_{ext}$(cm) | Density (g/cm$^3$) | Source Strength (n/s) |
|---|---|---|---|---|---|---|
| 1 | BF3 | $^{10}$B | 23.75 | 3.85 | 6.2485E-4 | --- |
| | | $^{11}$B | 1.25 | | | |
| | | $^{19}$F | 75.0 | | | |
| 2 | VACUUM | --- | --- | 7.5 | 0. | --- |
| 3 | FISSILE SOLUTION | $^{1}$H | 62.524489 | 19.895 | 1.162 | --- |
| | | $^{16}$O | 35.728280 | | | |
| | | $^{14}$N | 1.567612 | | | |
| | | $^{235}$U | 0.168450 | | | |
| | | $^{238}$U | 0.011169 | | | |
| 4 | SOURCE | $^{252}$CF | 100.0 | 19.896 | 1.0E-10 | 20000 |

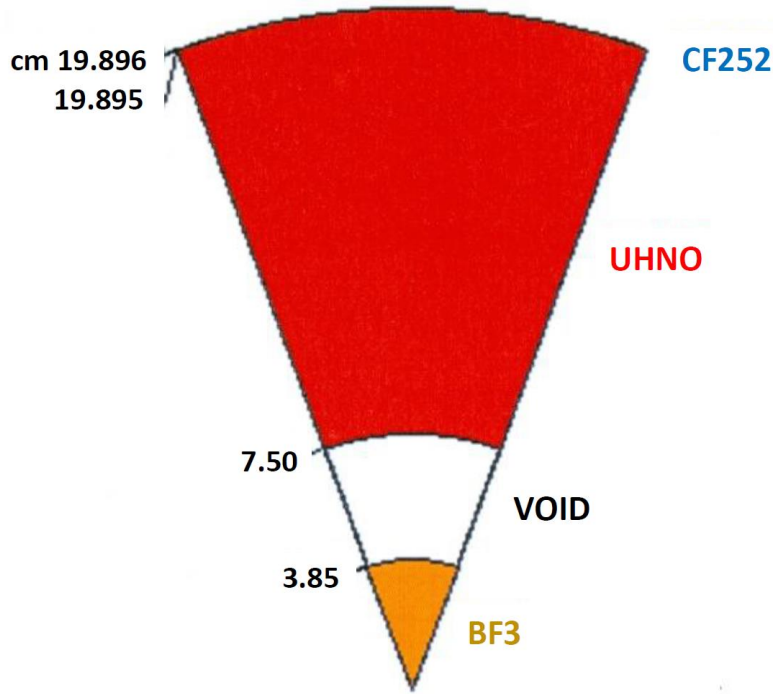

*Figure 1.* Test problem 1: A spherical fissile solution of uranyl nitrate with an inner BF3 detector and an external $^{252}$Cf spontaneous fission source.

IV.B.1.b. Calculation conditions

Deterministic calculations are performed using the following discretization:

- Space: 210 cells (50, 50, 100, 10 cells by materials starting from the center).
- Angle: $S_{16}$ equi-weight quadrature.
- Anisotropic scattering cross sections with 5 Legendre moments.
- Energy: 30 groups nuclear data based on the ENDF/B-VII evaluation, two multigroup files are used corresponding to prompt only induced fission neutrons and total (prompt plus delayed) fission neutron production.
- Time: variable time step between $10^{-6}$ and $10^{-3}$ s for time-dependent calculations.

Induced and spontaneous fission multiplicity probabilities are Terrell distributions[26]. For the spontaneous fission of $^{252}$Cf the parameters of the distribution are $\overline{\nu} = 3.76$ and $\sigma = 1.245$. The spontaneous fission source spectrum is the $^{252}$Cf Watt spectrum[25].

MCNP6.1 analog calculations are performed using ENDF/B-VII.1 nuclear data[27] and NPS=$10^7$ spontaneous fissions corresponding to a measurement time of 1880 seconds. A time list file is generated from the MCNP PTRAC file. Feynman moments and probability distributions of count numbers are obtained by processing the time list file. Using the source event number associated with each detection time, Feynman moments are calculated using correlated counts only[28]. The resulting Feynman curves are then more accurate and less fluctuating.

IV.B.1.c. Density Adjustment and Effective Source in PANDA Calculations

As the multiplicity counting parameters are very sensitive to system reactivity, the fissile density used for PANDA calculations is adjusted so that the estimated $k_{eff}$ is close to that given by MCNP6.1. The results are shown in table II. The density of the fissile solution used in PANDA is increased to 1.716 g/cm3 so that the $k_{eff}$ approaches the value obtained with MCNP6.1.

**Table II.** $k_{eff}$ calculation results.

|  | $k_{eff}$ | |
|---|---|---|
|  | nu prompt | nu total |
| PANDA (30g), ρ=1.602 g/cm³ | 0.94381 | 0.95165 |
| PANDA (30g), ρ=1.716 g/cm³ | 0.95358 | 0.96148 |
| MCNP6.1, ρ=1.602 g/cm3 | 0.95352 (σ=30. 10⁻⁵) | 0.96213 (σ=30. 10⁻⁵) |

When the time gate width is short compared with the decay time constants of the delayed neutron precursors, correlated counts come only from prompt neutrons. As a result, Feynman moments are calculated without considering delayed neutrons by using the nuclear data for prompt neutrons. However, delayed neutrons cannot be neglected when considering the average detection rate and the histogram of the number of counts. Because count numbers are proportional to source intensity, prompt neutron data can still be used if delayed neutrons are included in the source term.

Count rates calculated with ($R_T$) and without ($R_P$) delayed neutrons are presented in table III:

**Table III.** Count rates calculated with PANDA and MCNP6.1.

|  | PANDA | MCNP6.1 |
|---|---|---|
| $R_P$ (c/s) | 128.18 | 129.67 (σ=0.06) |
| $R_T$ (c/s) | 156.22 | 158.06 (σ=0.06) |

In order to take into account the delayed neutrons, PANDA calculations are performed using prompt neutron data and an effective neutron source given by:

$$S_{eff} = S \times \left(\frac{R_T}{R_P}\right)_{PANDA} = 24375 \text{ c/s} \qquad (95)$$

IV.B.1.d. Feynman Moments

In table IV we present the results of Feynman moment calculated with PANDA and MCNP6.1. The asymptotic values of the second and third Feynman moments are computed with PANDA in the stationary mode. The Feynman moments for a time gate width T=20ms are obtained with PANDA in the time dependent mode. The PANDA results are compared to those obtained with MCNP6.1. These results show a good agreement between deterministic and Monte Carlo calculations. We also verify that the delayed neutrons have no significant effect on the Feynman moments.

**Table IV.** Feynman moments calculated with PANDA and MCNP6.1.

|  | PANDA | MCNP6.1 nu prompt | MCNP6.1 nu total |
|---|---|---|---|
| $Y_2$(T=20ms) | 1.2115 | 1.19 (σ=0.01) | 1.19 (σ=0.01) |
| $Y_{2\infty}$ | 1.2638 |  |  |
| $Y_3$(T=20ms) | 4.2815 | 4.17 (σ=0.12) | 4.22 (σ=0.12) |
| $Y_{3\infty}$ | 4.6642 |  |  |

Several points on the Feynman curves are calculated with PANDA and plotted against the MCNP6.1 curves (figure 2 and 3). We observe an excellent overall agreement.

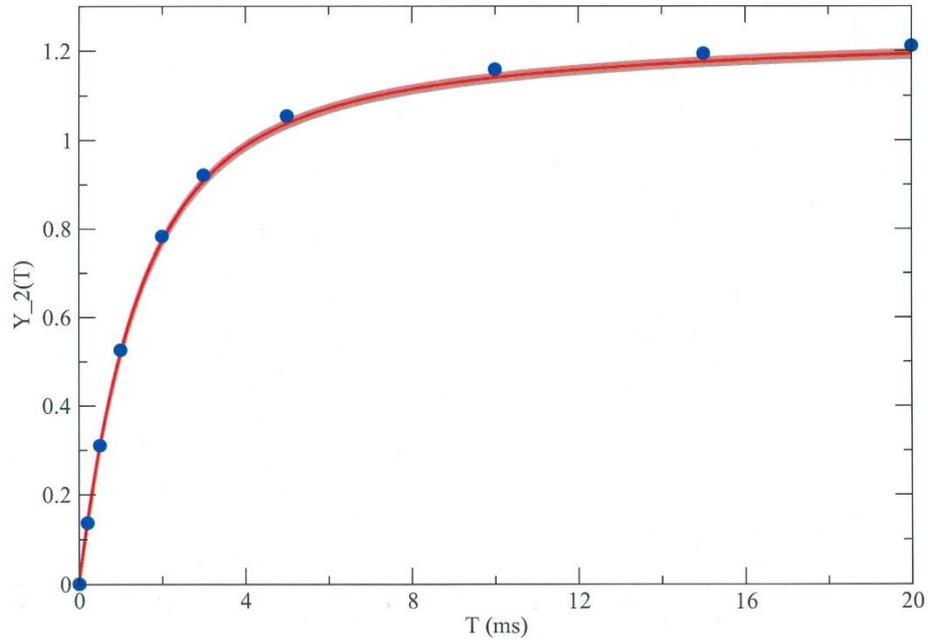

*Figure 2.* Second-order Feynman function calculated using MCNP6.1 (red curve with one-sigma uncertainties in brown). Blue points correspond to time-dependent PANDA calculations.

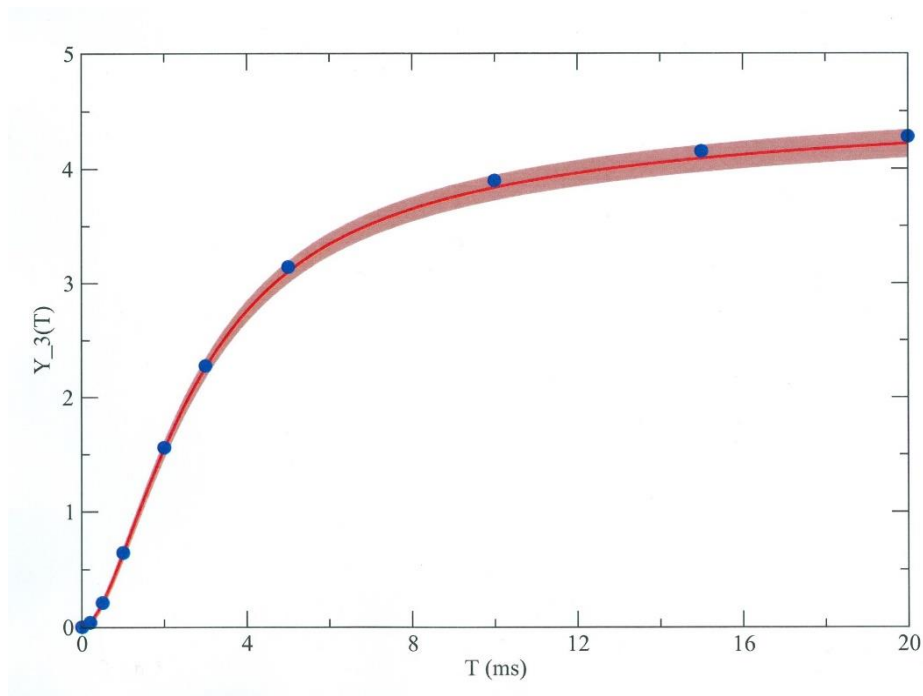

*Figure 3.* Third-order Feynman function calculated using MCNP6.1 (red curve with one-sigma uncertainties in brown). Blue points correspond to time-dependent PANDA calculations.

IV.B.1.e. Neutron Count Probability Distributions

The static probability distributions for a time gate width T=100 ms calculated with MCNP6.1 and with PANDA in stationary mode are presented in figure 4. Because the time gate is sufficiently large, the distribution has reached its asymptotic value and the static calculation agree with the MCNP6.1 distribution.

The effect of the time dependence is illustrated in figure 5. For a time gate width T=5ms, the distribution has not yet reached its asymptotic shape, however the time-dependent PANDA calculation is perfectly consistent with the MCNP6.1 result.

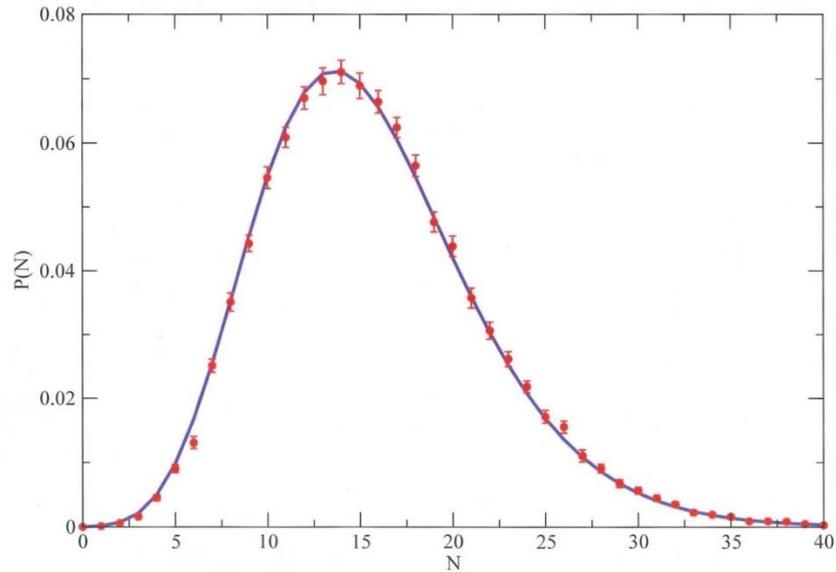

*Figure 4.* MCNP6.1 count number probability distribution for a time gate width T=100 ms (in red) compared to the distribution obtained with a static PANDA calculation (in blue).

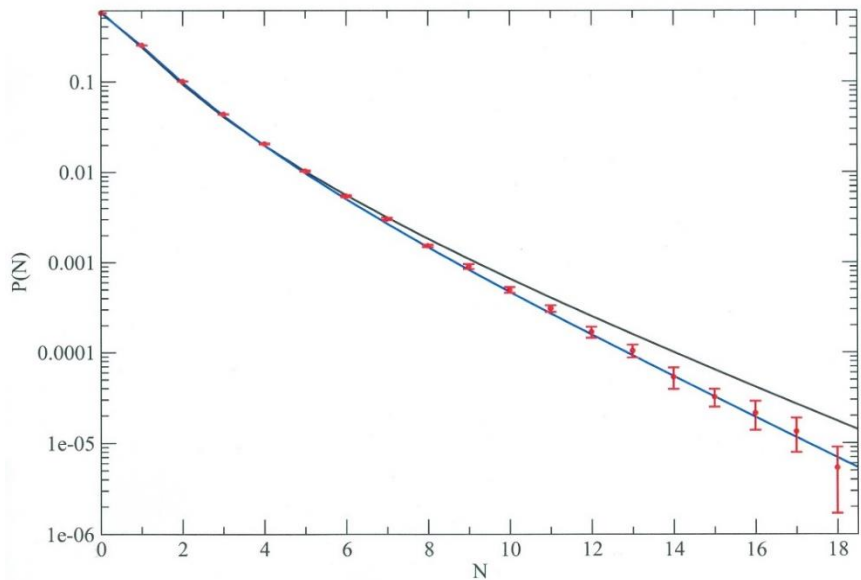

*Figure 5.* MCNP6.1 count number probability distribution for a time gate width T=5ms calculated using MCNP6.1 (red) compared to the distribution obtained with a time-dependent PANDA calculation (blue) and a static PANDA calculation (black).

*IV.B.2. Second test problem*

A second test problem is designed to study the effect of truncating, to order N lower than the maximum I, the sequence of moments for induced and spontaneous fission multiplicities:

$$v_i = \hat{v}_i = 0 \text{ for } i \geq N$$

This means that the fission trees selected have no branching nodes giving birth to more than N sub-chains ultimately contributing to the detection. This is referred to as the N-forked fission branching approximation[29].

IV.B.2.a. Test problem description

The 1D spherical test problem is illustrated in Fig. 6. It consists of a highly enriched uranyl nitrate solution with centrally placed $^{252}$Cf spontaneous fission source. The number of detections is a fraction $\varepsilon_L$ of the total neutron leakage during a time gate width T.

The dimensions, density and atomic composition of the fissile solution are shown in Table V. The other characteristics of the problem are:

- Intensity of the neutron source: $S = 1000$ n/s
- Leakage efficiency: $\varepsilon_L = 10\%$
- Detection time gate: $T = 10$ ms

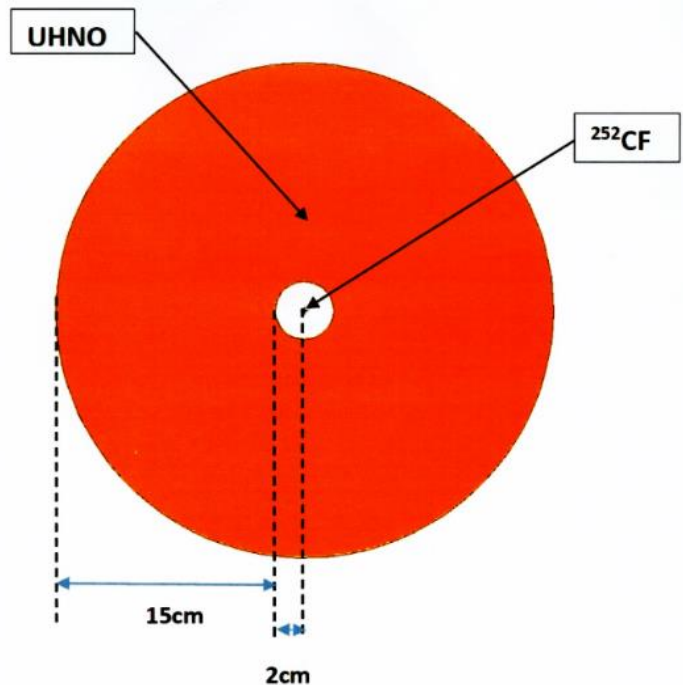

*Figure 6.* Test problem 2: A highly enriched uranyl nitrate solution with centrally placed $^{252}$Cf spontaneous fission source.

**Table V.** Characteristics of test problem 2.

| Shell # | Material | Isotope | at% | $R_{ext}$(cm) | Density (g/cm$^3$) |
|---|---|---|---|---|---|
| 1 | SOURCE | $^{252}$Cf | 100 | 0.001 | 1.E-10 |
| 2 | VACUUM | --- | --- | 2 | 0. |
| 3 | FISSILE SOLUTION | $^{1}$H | 62.524489 | 17 | 1.160 |
| | | $^{16}$O16 | 35.728280 | | |
| | | $^{14}$N14 | 1.567612 | | |
| | | $^{235}$U235 | 0.168450 | | |
| | | $^{238}$U238 | 0.011169 | | |

IV.B.2.b. Calculation conditions

The deterministic computations are performed using PANDA code with the following discretization:

- Space: 160 cells (10,50,100 cells by material starting from the center)
- Angle: $S_{16}$ equi-weight quadrature
- Energy: 30 groups nuclear data from the ENDF/B-VI evaluation, using prompt fission neutron multiplicity
- Anisotropic scattering cross sections with 5 Legendre moments
- Stationary calculation

Fission multiplicity probability distributions and source spectrum are the same as in problem 1.

IV.B.2.c. Results

The calculated prompt multiplication factor is $k_{eff}$ = 0.90399.

Binomial moments and counting probabilities are calculated using increasing levels of the N-forked branching approximation. The results are reported in Table VI.

**Table VI.** Binomial moments results using N-forked branching approximation. Results in bold red are exact and results in black are approximations

|       | N=1     | N=2      | N=3      | N=4      | N=5      |
|-------|---------|----------|----------|----------|----------|
| $M_1$ | **6.9747** | **6.9747** | **6.9747** | **6.9747** | **6.9747** |
| $M_2$ | 24.323  | **68.649** | **68.649** | **68.649** | **68.649** |
| $M_3$ | 56.549  | 800.69   | **815.35** | **815.35** | **815.35** |
| $M_4$ | 98.603  | 10462    | 10902    | **10905**  | **10905**  |
| $M_5$ | 137.545 | 147444.4 | 157534.1 | 157641.2 | **157641.7** |

When all fission multiplicity moments of order greater than one are set to zero (column N=1 in Table VI), there is no correlation and the count distribution obeys Poisson's law. In this case, we verify that the binomial moments of the count number distribution are:

$$M_n = \frac{(M_1)^n}{n!} \tag{96}$$

We also check that, when using the N-forked branching approximation, the first N moments of the counting distribution are exact (shown in bold red in table V), and moments of order greater than N are approximations.

The probability distributions calculated using the two-, three- and eight-forked branching approximations are depicted in Figure 7. In this example, the two-forked (quadratic) approximation is not accurate, and we observe that a good approximation is obtained using at least three moments (cubic approximation) of the fission multiplicity distributions. We also note that, unlike binomial moments, probabilities depend on all moments of the fission multiplicity.

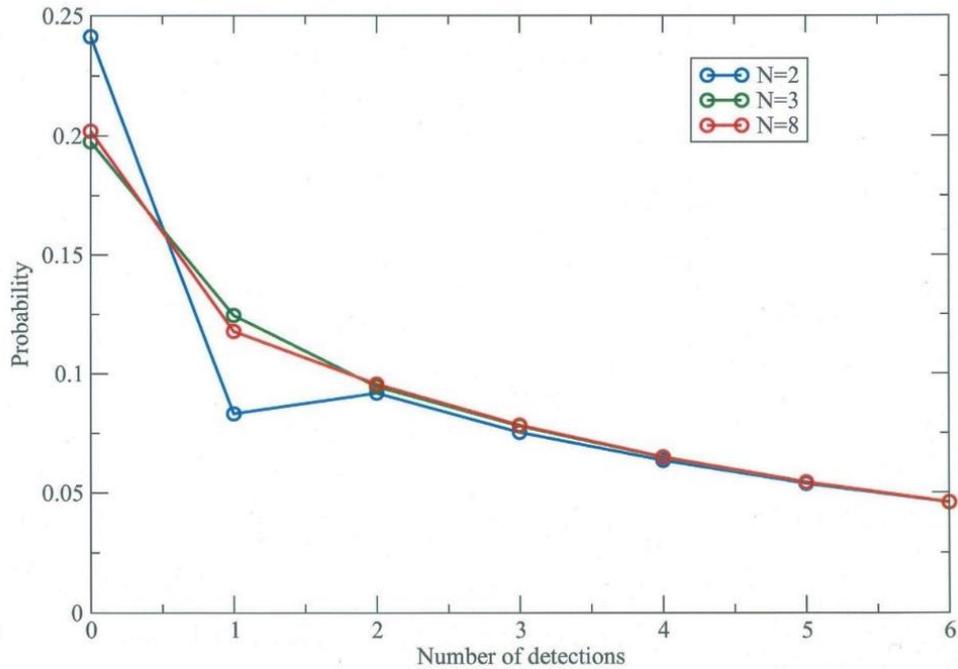

*Figure 7.* Count number probability distributions for different level of branching approximation (N=2, 3 and 8).

**V. CONCLUSION**

We have investigated the probability and moment equations for the number of neutron detections induced by a single initial neutron, and then by a neutron source. These equations have been derived, based on the methodology of Pàl and Bell, using a simplified point model. It was observed that the moments and probability equations are similar, with the exception of the zero-order moment, which is just equal to one, while the probability of the zero event is the solution of a non-linear equation. In both cases, we have obtained a set of differential equations coupled by their source term. We showed that these equations can be solved successively up to an arbitrary order using recursive formulas. It was also pointed out that the same equations can be used for neutron leakage or population number distribution by simply adapting by the adjoint boundary and final time conditions.

In the second place, the point model problem has been generalized to take into account the complete neutron phase space. The ordinary differential equations of the point model have become conventional linear adjoint transport equations, with the exception of the zero event probability, which has a non-linear part. It is therefore easy to adapt a deterministic transport code to solve these equations, and even the non-linearity can be solved during source iteration. In this work, the implementation was carried out using the deterministic discrete ordinates code PANDA.

We then presented two 1D spherical test cases to verify the applicability of the updated PANDA code. In the first test case, the PANDA code was checked against the Monte Carlo reference code MCNP6.1. A good agreement was found for static and time-dependent calculations of the second and third Feynman moments and for the probability distributions of the detection numbers. However, as the multigroup nuclear data file used in PANDA calculations was not sufficiently converged regarding the group number, it was decided to adjust the fissile solution density in PANDA to have the same effective multiplication coefficient value as with MCNP6.1. For the count distributions, delayed neutrons were taken into account by adjusting the source intensity.

In the second test case, we have considered the number of detected neutrons escaping through the outer boundary. In particular, we studied the N-forked branching approximation, which consists in truncating the number of multiplicity moments used in the induced and spontaneous fission terms. We have verified that the Nth-order binomial moment is calculated exactly using the first N lower-order moments of the fission multiplicity. Furthermore, when considering the probability distribution of the detection number, we found that a good approximation was obtained using at least three moments (cubic approximation) of the fission multiplicity distributions.

In terms of computational efficiency, we observed that the stationary calculation of the first three moments in 1D is fast, typically a few seconds for a sequential calculation. However, time dependent calculations of the probability distributions can become costly when the number of transport equations to be solved is large. In further work, we intend to apply the code to neutron population problems and consider 2D test problems with parallelization.